\documentclass[final,leqno,onefignum,onetabnum]{siamltex1213}

\usepackage{color,amsmath,epsfig,epsf}
\usepackage{fullpage}
\usepackage{url}
\usepackage{listings}

\definecolor{gray}{rgb}       {0.8,0.8,0.8}
\definecolor{light-blue}{rgb} {0.8,0.8,1.0}
\definecolor{light-green}{rgb}{0.8,1.0,0.8}
\definecolor{light-red}{rgb}  {1.0,0.9,0.9}

\newcommand{\commentout}[1]{}

\newcommand{\Cplusplus}{{\rmfamily C\raise.22ex\hbox{\small ++} }}

\begin{document}

\title{BoxLib with Tiling: An AMR Software Framework
\thanks{This work was supported by the SciDAC Program and
the Exascale Co-Design Program of the DOE Office of Advanced
Scientific Computing Research under the U.S. Department of Energy
under contract DE-AC02-05CH11231.  This research used resources of the
National Energy Research Scientific Computing Center, which is
supported by the Office of Science of the U.S. Department of Energy
under Contract No. DE-AC02-05CH11231. Dr. Unat is supported by the Marie Sklodowska Curie 
Reintegration Grant by the European Commission, and by the Tubitak Grant No: 215E285.
We would like to acknowledge and
thank Brian Friesen for running some of the performance 
tests and John Bell for many useful discussions concerning tiling. 
}}

\author{Weiqun Zhang   \footnotemark[2]
\and    Ann Almgren    \footnotemark[2]
\and    Marcus Day     \footnotemark[2]
\and    Tan Nguyen     \footnotemark[3]
\and    John Shalf     \footnotemark[4]
\and    Didem Unat     \footnotemark[5]}

\renewcommand{\thefootnote}{\fnsymbol{footnote}}

\footnotetext[2]{Center for Computational Sciences and Engineering,
                 Lawrence Berkeley National Laboratory,
                 One Cyclotron Road MS50A-1148,
                 Berkeley, CA 94720, USA.
                 https://ccse.lbl.gov/
                 Email: WeiqunZhang@lbl.gov, ASAlmgren@lbl.gov, MSDay@lbl.gov}
\footnotetext[3]{Computer Architecture Group,
                 Lawrence Berkeley National Laboratory,
                 One Cyclotron Road MS50A-1148,
                 Berkeley, CA 94720, USA.
                 crd.lbl.gov/departments/computer-science/computer-architecture/staff/tan-thanh-nhat-nguyen,
                 Email: TanNguyen@lbl.gov}
\footnotetext[4]{Computer Science Department,
                 Lawrence Berkeley National Laboratory,
                 One Cyclotron Road MS50A-1148,
                 Berkeley, CA 94720, USA.
                 crd.lbl.gov/departments/computer-science/,
                 Email: JShalf@lbl.gov}
\footnotetext[5]{Computer Science and Engineering Department,
                 Ko\c{c} University,
                 Rumelifeneri Yolu, 
                 Sariyer, Istanbul, Turkey 34450
		http://parcorelab.ku.edu.tr
                 Email: dunat@ku.edu.tr}

\renewcommand{\thefootnote}{\arabic{footnote}}

\pagestyle{myheadings}
\thispagestyle{plain}

\maketitle

\begin{center}
{\bf Abstract}
\end{center}

In this paper we introduce a block-structured adaptive mesh refinement (AMR)
software framework that incorporates tiling, a well-known loop transformation. 
Because the multiscale, multiphysics codes built in BoxLib
are designed to solve complex systems at high resolution, performance
on current and next generation architectures is essential.
With the expectation of many more cores per node on next generation architectures,
the ability to effectively utilize threads within a node is essential, and 
the current model for parallelization will not be sufficient.
We describe a new version of BoxLib in which the tiling constructs are embedded 
so that BoxLib-based applications can easily realize expected performance gains
without extra effort on the part of the application developer. We also discuss
a path forward to enable future versions of BoxLib to take advantage of NUMA-aware optimizations 
using the TiDA portable library.

\section{Introduction}

BoxLib is a mature, publicly available, software framework for
building massively parallel block-structured adaptive mesh refinement
(AMR) applications~\cite{BoxLib}.  It is one of a number of current publicly
available AMR frameworks; see \cite{Dubeyetal:2014}
for an overview.  Numerous application codes are based on these
frameworks and are too many to list here, but sample research codes in
use today that are based on BoxLib include MAESTRO
\cite{ABRZ:I,ABRZ:II,ABNZ:III,ZABNW:IV,MAESTRO:Multilevel} and CASTRO
\cite{CASTRO_I,CASTRO_II,CASTRO_III} for astrophysical simulations,
Nyx \cite{Nyx,Nyx2} for cosmological simulations, LMC \cite{DayBell:2000,LMC2,LMC3}
for low Mach number combustion simulations, SMC \cite{SMC} for
compressible combustion simulations, as well as codes for moist
atmospheric physics \cite{CASTROmoist,MAESTROmoist} and subsurface
flow \cite{Pau:2012}.

BoxLib contains extensive software support for explicit and implicit 
grid-based operations as well as particle-mesh operation on adaptive hierarchical meshes. 
Multilevel multigrid solvers are included for cell-based and node-based data.  
Multiple time-subcyling modes are supported for adaptive mesh simulations.
Most current BoxLib application codes evolve multiscale, multiphysics
systems in time by solving systems of partial differential equations (PDEs) 
often accompanied by constraints.
Because these codes are designed to solve complex
systems at high resolution, performance on current and next generation
architectures is always of paramount importance.

BoxLib uses a hybrid MPI/OpenMP approach for parallelization, and the
OpenMP parallelism has traditionally been expressed in the individual 
loops over cells.  BoxLib application codes have demonstrated
good scaling behavior on up to 100,000 cores on current multi-core architectures 
(see e.g., \cite{CASTRO_I,Malone_etal:2014}). 
However, with the current trends in system design, the next generation
of high-performance computing systems 
will have node architectures based on many-core processors and non-uniform memory access (NUMA) designs. 
Due to the irregularity of AMR algorithms, a fine-grained loop-level threading approach 
is not expected to provide efficient parallelism on a node with hundreds or thousands of threads.  
In addition, reducing the cost of data movement is expected be crucial for performance \cite{padal-report}.
In response to these architectural challenges, we are adopting a new
programming model, {\em tiling}, in BoxLib to expose additional parallelism
and optimize data access behavior.

Tiling is a well known loop transformation that has been proven to be useful in
enhancing data locality and parallelism (e.g., \cite{Wolfe:1989,Rivera:2000}).  
Tiling reduces the working set size so that the working set of a thread can fit 
into cache, reducing the number of cache misses.  This decreases the memory traffic, 
resulting in improved performance.
Tiling can also expose more parallelism by multi-dimensionally partitioning data or 
iteration space. Previously we added tiling to the BoxLib based SMC code \cite{SMC} manually.  
However, manual tiling of an application code is both labor intensive and error prone because 
there is no standard language construct for tiling in any of the general-purpose programming languages. 
In this paper, we introduce a new version of BoxLib that embeds tiling
in the shared framework rather than in each application code.  By supporting tiling 
in the Boxlib framework itself, application developers using BoxLib do not have to
recode any of the tiling constructs.

In Section~\ref{sec:tilingoverview}, we present a brief introduction to tiling.
In Section~\ref{sec:boxlib}, we give a brief overview of BoxLib and key
abstractions in the pre-tiling version are presented.  We then present
the new version of BoxLib with tiling and discuss the threading approach. Several performance results are
shown in Section~\ref{sec:performance}, demonstrating the improved performance of 
the tiling approach.  In Section~\ref{sec:regional}, we discuss our plans for 
further optimization of BoxLib using the TiDA portable library \cite{TiDA,TiDA-ISC}.  We summarize the results
of this paper in the final section.


\section{Tiling Overview and Related Work}
\label{sec:tilingoverview}

Modern systems contain complex memory hierarchies with multiple levels
of caches and multiple NUMA nodes. 
It is essential that a software framework like BoxLib adapt
to the increasing number of cores and possibly increasing number of NUMA domains
in order to be able to take advantage of the current and future architectures.
Tiling, also known as cache blocking, is a well known program
transformation that has proven to be useful in improving data locality
and parallelism.  Tiling eliminates some of the memory traffic by reducing 
the size of the working set so that there are fewer cache misses. 
In addition, it enables high degrees of parallelism through partitioning iteration space into
independent units of execution.  At the language level, there is no
standard programming support to express tiling information in the
program. The programmer has to manually tile individual loops, which can be very
labor-intensive and error-prone for large codes like the BoxLib-based 
AMR applications.
The most common way to implement tiling is to partition a nested loop by introducing
{\em tiling loops} that iterate over tiles and {\em element loops} that iterate over 
the data elements within a tile. 

There is a long list of literature on iteration space tiling
to move the burden from the programmer to the compiler
~\cite{Wolfe:1989, HTA, Rivera:2000:tiling, 
Carr:1992:CBN, Goumas:2004:APC, Song:1999:NTT}. 
Compiler methods rely on static loop transformations~\cite{ICS11:Mint}, usually in the form 
of source-to-source translation. There is only limited support for tiling 
in commercial compilers because of the code
generation complexity. The complexity emerges primarily from the semantics of 
existing programming languages (e.g. C, C++, or Fortran) and 
the inability in current programming environments to express essential information 
required for effective performance analysis and safe code transformations to auto-generate tiled code.
It is particularly difficult for AMR codes,
where crucial information needed to optimize data locality is only available 
at runtime. 
Our approach here is to decouple tiling from the individual loops and instead represent it 
at the data structure level within the BoxLib framework. 

HTA (Hierarchically Tiled Arrays)~\cite{HTA} offers data structures for describing a hierarchy and topology 
of tiles where computation and communication are represented by overloaded array operations. 
Extending array notations hides many details from the programmer but eventually 
creates performance problems because of the excessive use of temporary arrays 
or frequent data layout transformations. Adopting HTA in BoxLib would have meant
extensive rewriting of BoxLib and BoxLib-based application codes.
TiDA~\cite{TiDA,TiDA-ISC}, on the other hand, is a library that allows parameterised 
data layout and tiling but does not require restructuring of the original software framework. 
We are motivated here by TiDA's approach to tiling and have integrated many features of TiDA into the BoxLib framework. 

\begin{figure}
\centering
\includegraphics[width=3.5in]{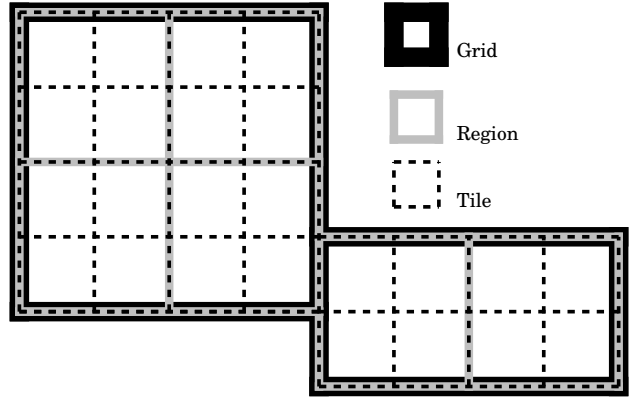}
\caption{\label{fig:GRT} Grid, Region and Tile. In this example of
  regional tiling, there are two grids.  The grid on the left contains
  four regions, and the grid on the right contains two regions.  Each
  region is split into four logical tiles in iteration space.  The
  floating point data are allocated contiguously in each region, thus
  the data within a grid are not in one contiguous block.  Note that
  logical tiles do not affect the data layout in memory.}
\end{figure}

TiDA introduces two concepts: {\em regional} and {\em logical}
tiling. Figure~\ref{fig:GRT} shows an example of regional tiling.  
We denote a rectangular domain in index space as a {\em grid};
each grid is tiled into multiple {\it regions} with each region's 
floating point data allocated contiguously in memory.  Each region is further split
into {\em logical tiles} in iteration space; the logical tile size
can be changed dynamically on a loop-by-loop basis.  It should be
emphasized that logical tiles exist only in the iteration space and
do not alter the floating point data layout in memory.  A tiling
approach based on the hierarchy of grid, region and tile is called
regional tiling.  Logical tiling is a special case of regional tiling
in which each grid contains only one region.  Thus the data on a grid
in logical tiling are contiguous in memory, whereas in regional tiling
they are allocated in multiple contiguous blocks.  Regional tiling
is intended to address the NUMA nodes and regional coherence domains,
and is more general than logical tiling; however it requires data
structure changes because existing AMR frameworks usually allocate one
contiguous block of memory for the data on each grid.  In contrast,
logical tiling is easily adoptable by the underlying framework,
because it requires minimal bookkeeping and incurs little overhead on
existing codes.

\section{BoxLib}
\label{sec:boxlib}
The BoxLib software framework includes both a \Cplusplus framework that can call
Fortran subroutines, and an entirely Fortran framework.  Both
frameworks have been modified to use tiling, but we will present
the illustrative examples in \Cplusplus only for simplicity.

\begin{figure}[h]
\centering
\includegraphics[width=2in]{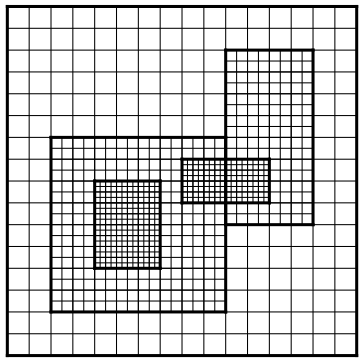}
\caption{\label{fig:grid_cartoon} Cartoon of AMR grids with two levels
  of factor 2 refinement.  The coarsest grid covers the domain with
  $16^2$ cells.  Bold lines represent grid boundaries.  The two
  intermediate resolution grids are at level 1 and the cells are a
  factor of two finer than those at level 0.  The two finest grids are
  at level 2 and the cells are a factor of two finer than the level 1
  cells.  Note that there is no direct parent-child connection. The
  data for each grid are contained in an object called
    FArrayBox (see section~\ref{sec:boxlib0}).  In this example,
  there are 1, 2 and 2 FArrayBox objects on level 0, 1, and 2,
  respectively.  FArrayBox objects on each level are organized
  into an object called MultiFab (Multiple FArrayBox).  }
\end{figure}

The data in BoxLib is defined on a nested hierarchy of logically
rectangular grids.  Recall we use the term {\it grid} to refer to a
rectangular domain in index space; in the context of block-structured
AMR a {\it level} is composed of a union of (in this case
non-intersecting) grids that share the same mesh (or cell)
spacing.  The ratio of mesh spacings between adjacent levels is
typically 2 or 4.  Figure~\ref{fig:grid_cartoon} shows a cartoon of
AMR grids in two dimensions with two levels of refinement.  One aspect
that distinguishes BoxLib from frameworks such as FLASH \cite{FLASH}
is that a fine grid does not necessarily have a unique parent grid.
This allows the organization of data into relatively large aggregate
grids and as a result amortizes the cost incurred by the irregular
nature of adaptive meshes. 

The changes to BoxLib described in this paper focus on data
locality and parallelism at each level of refinement independently,
thus for the rest of the paper we will confine ourselves to the
geometry of a single level.  However, the fact that a tiling strategy
must work in the context of complex multiphysics applications on
adaptive grid hierarchies dictates three necessary features.  First,
the strategy must work for a union of grids that are not necessarily
of equal size and shape, and that do not necessarily span the entire
rectangular domain. Second, the strategy must be such that the tile
size can be modified depending on the nature of the loop, as different
parts of the algorithm may have very different computational and
communication demands.  Finally, the tiling strategy must be
sufficiently lightweight to adapt to the frequently changing grid
structure at all levels but the coarsest as the simulation evolves.

\subsection{BoxLib Without Tiling}
\label{sec:boxlib0}

Before we discuss the tiling strategy, we describe 
the data structures and methods for operating on the data on a level
in BoxLib without tiling.  In BoxLib, a {\it Box} is the data structure for
a rectangular domain in index space, and a three-dimensional Box can be
represented by six integers. Each grid is represented by the Box type.  
A {\it FArrayBox} is the data structure that holds the floating point data 
on a single Box.  The data can have multiple components (such as density and velocity), 
and are stored internally as a one-dimensional (1D) array allocated with \Cplusplus's
{\tt new} operator, therefore are contiguous in memory.  The index
space information in FArrayBox can be used to reshape the 1D
array into a multi-dimensional array to be used in a Fortran
subroutine.  We note that while the grids themselves are 
non-intersecting boxes, the data in an FArrayBox may be defined
on a Box larger than the grid if ghost (or halo) cells are needed.
A {\it MultiFab} is the parallel data structure containing multiple
FArrayBoxes which holds data on the multiple grids of an AMR
level (see figure~\ref{fig:grid_cartoon}).  The data in a 
MultiFab are distributed among MPI processes.
Thus, on each MPI process the MultiFab contains only the FArrayBox 
objects owned by this process, and the process operates only on its local data.  
For operations that require data owned by other MPI processes, ghost cells
are filled via MPI communications.  To reduce the
  latency and use the network bandwidth more efficiently, the
communication messages are aggregated.  For the purpose of communication
within each level and between levels, each MPI process has a copy of
all the grids. 

Many common operations in BoxLib can be performed at the abstract
level of MultiFab with the data and implementation details
hidden from the application developers.  A few examples are given below.
\begin{lstlisting}
   // Return the maximum value contained in component 0 of the MultiFab mf
   max_val = mf.max(0);

   // Fill the ghost cells of each FArrayBox in the MultiFab mf
   mf.FillBoundary();
\end{lstlisting}
To operate on the data in more application-specific ways, BoxLib supplies iterators,
called {\tt MFIter}s, over the FArrayBoxes in a MultiFab.   The application developer provides
a Fortran subroutine that performs the operations; each subroutine operates
on one grid's worth of data at a time.
For example, 
\begin{lstlisting}
  for (MFIter mfi(mf); mfi.isValid(); ++mfi) // Loop over grids
  {
    // Define vbox to be the grid associated with this iteration.
    // This "valid" region will be used to define the extent of the data
    // that the subroutine will operate on.
    const Box& vbox = mfi.validbox(); 

    // Get a reference to the FArrayBox so that we can access 
    // both the data and the size of the FArrayBox.
    FArrayBox& fab = mf[mfi];  

    // Define the double* pointer to the data of this FArrayBox.
    double* a = fab.dataPtr();

    // Define abox as the Box on which the data in the FArrayBox is defined.
    // Note that "abox" includes ghost cells (if there are any), and is thus
    // larger than or equal to "vbox".
    const Box& abox = fab.box();

    // We can now pass the information to a Fortran routine,
    // which treats the double* as a multi-dimensional array 
    // with dimensions specified by the information in "abox".
    // We will also pass the index information in "vbox", 
    // which specifies our "work" region.
    // Functions loVect() and hiVect() return the lower and upper
    // indices of a box, respectively.
    f(vbox.loVect(), vbox.hiVect(), a, abox.loVect(), abox.hiVect());
  }
\end{lstlisting}
The Fortran subroutine might look as shown here:
\begin{lstlisting}
  subroutine f(lo, hi, a, alo, ahi)
    integer, intent(in) :: lo(3), hi(3), alo(3), ahi(3)
    double precision, intent(inout) :: a(alo(1):ahi(1),alo(2):ahi(2),alo(3):ahi(3))
    integer :: i, j, k
    !$OMP PARALLEL DO private(i,j,k)
    do     k = lo(3), hi(3)
      do   j = lo(2), hi(2)
        do i = lo(1), hi(1)
           ! ...
        end do
      end do
    end do
    !$OMP END PARALLEL DO
  end subroutine f
\end{lstlisting}
Here, for the sake of simplicity, we have omitted the code that facilitates
calling Fortran subroutines from \Cplusplus.  In BoxLib, this is
usually handled by C preprocessor macros; one can also use the
ISO\_C\_BINDING module in Fortran 2003.  In this example, OpenMP is
used on the loop level for work sharing among threads.

\subsection{BoxLib With Logical Tiling}
\label{sec:boxlib1}

In the new version of BoxLib\footnote{BoxLib is under active
development and all updates are publicly available through git as
soon as they are tested and committed.  By ``new version'', we do not refer to any particular
version number.}, we introduce the capability for {\it logical tiling} as
defined earlier.  For backward compatibility the default is to have
tiles the same size as (or larger than) the grids so that tiling is
effectively turned off.  In our logical tiling approach, each grid is
now logically split into tiles (both grids and tiles are of type
Box), and the modified looping construct is completely incorporated
into the MFIter iterator which now loops over each tile in each
grid rather than just over the grids themselves.  With logical tiling,
the data layout is unchanged; just the access pattern to the data is
changed.  An example of using tiling is shown below.

\begin{lstlisting}
  bool tiling = true;
  for (MFIter mfi(mf,tiling); mfi.isValid(); ++mfi) // Loop over tiles
  {
    // Define the tile of this iteration
    // This tile, rather than the grid that the tile is a part of,
    // will be used to define the extent of the data
    // that the subroutine will operate on.
    const Box& tbox = mfi.tilebox(); 

    // Get a reference to the FArrayBox so that we can access
    // both the data and the size of the FArrayBox; the FArrayBox
    // is unchanged by using tiling
    FArrayBox& fab = mf[mfi];  

    // Define the double* pointer to the data of this FArrayBox.
    // The dataPtr of the FArrayBox is unchanged by using tiling.
    double* a = fab.dataPtr();

    // Define abox as the Box on which the data in the FArrayBox is defined.
    // This is also unchanged by using tiling.
    const Box& abox = fab.box();

    // We can now pass the information to a Fortran routine.
    // We now pass the index information in "tbox" to specify the work region.
    f(tbox.loVect(), tbox.hiVect(), a, abox.loVect(), abox.hiVect());
  }
\end{lstlisting}
In this example, the loop controlled by MFIter can be regarded
as a nested loop collapsed into one iteration space, where the nested
loop is composed of an outer loop of grids and an inner loop of tiles
in a grid.  Note that the code is almost identical to the one in
\S~\ref{sec:boxlib0}; the Fortran subroutine and the data pointer
passed to it are unchanged, but the index space on which the subroutine
operates is now defined by the tile rather than the grid.
Loop tiling is achieved without any modification in the kernel function f.  
This is important to application developers using BoxLib since
typically the kernels are application-specific.
Moreover, any local arrays in the kernel function
can now be the size of the tile rather than the potentially much larger grid, 
thus the working set size is further reduced, resulting in better data access patterns.

The MFIter class has several constructors:
\begin{lstlisting}
  MFIter(const MultiFab& mf);
  MFIter(const MultiFab& mf, bool do_tiling);
  MFIter(const MultiFab& mf, const int tilesize[]);
\end{lstlisting}
In the first version, each grid has one tile and the tiling is
effectively turned off.  If the second version is used, the default
tile size (that can be specified at runtime) is used.  The third version
provides more control over the tile size by passing an integer array
specifying the tile size in each spatial dimension.  This allows applications
to use different tile sizes in different parts of the algorithm,
which is essential for maximizing performance
in a complex multiphysics code.
The new 
{\tt tilebox} member function of MFIter returns a box for the
current tile of the iteration.  Additional MFIter member functions 
that are useful for building application codes include
\begin{lstlisting}
 Box growntilebox(int nghost) const;
 Box nodaltilebox(int direction) const;
\end{lstlisting}
The function {\tt growntilebox} returns a grown tile box that includes
ghost cells at grid boundaries only; therefore the returned
boxes originating from the same grid are still non-overlapping.  This
is useful when the work region includes ghost cells.  The function 
{\tt nodaltilebox} returns non-overlapping face-type boxes for tiles,
with the face specified by the integer argument.  This function is
particularly useful for finite-volume methods where face-based fluxes are
often used.

\subsection{Threading in BoxLib}
\label{sec:boxlib2}

BoxLib uses OpenMP for threading.  As seen in section~\ref{sec:boxlib0}, 
BoxLib traditionally used {\tt OMP DO} in a fine-grained loop-level threading approach.
Tiling provides the opportunity for a coarse-grained approach for
threading,  which enables the removal of OpenMP constructs from the individual Fortran routines.
For example, OpenMP parallelism at the tile level can now be achieved by adding OpenMP pragmas
around the MFIter loops:
\begin{lstlisting}
  #pragma omp parallel
  for (MFIter mfi(mf,true); mfi.isValid(); ++mfi) // Loop over tiles
  {
    const Box& tbox = mfi.tilebox(); 
    FArrayBox& fab = mf[mfi];  
    double* a = fab.dataPtr();
    const Box& abox = fab.box();
    f(tbox.loVect(), tbox.hiVect(), a, abox.loVect(), abox.hiVect());
  }
\end{lstlisting}
Here {\tt mfi} is a thread private MFIter object because it is
created inside an OpenMP parallel region.  Because an OpenMP parallel
region has already started when function f is called, we
remove the OpenMP directives in function f unless nested OpenMP
parallelism is desired.

When MFIter detects that it is inside an OpenMP parallel region as in
the code above, the iterations over tiles are shared among OpenMP 
threads.  In our current implementation, the work scheduling policy
is similar to OpenMP's static scheduling.  Let us consider an
example.  Suppose there are four grids of different sizes at a level,
and the tile size is such that the first and fourth grids are each
divided into 4 logical tiles, while the second and third are divided
into 2 logical tiles each.  In the tiling version, the loop body will
be run 12 times in total, as opposed to 4 times in the untiled version.  
Assuming four threads are used, thread 0 will work on 3
tiles from the first grid, thread 1 on 1 tile from the first grid and 2
tiles from the second grid, and so forth.  

In addition to the benefits of tiling itself,
this coarse-grained tile-level threading approach has several
advantages over the fine-grained loop-level approach shown in
Section~\ref{sec:boxlib0}.
First, when individual grids are small (such as often occurs after multiple
coarsenings near the bottom of a multigrid V-cycle in a multigrid solver), 
there can still be enough parallelism for the tile-level
approach even though there might not have been enough without tiling.  
Second, the tile-level approach has removed the need to use OpenMP directives in the 
application-specific Fortran subroutines in most cases, which reduces the chance of
application-specific OpenMP coding errors.  
Finally, it reduces thread overhead and synchronization
points because there is only one OpenMP parallel region as opposed to
many parallel regions in the fine-grained loop-level approach.

There could be an even coarser grained grid-level parallelism
approach, in which the operations on each grid are performed by only one
thread and different threads operate on different grids. 
This grid-level approach is less desirable than the tile-level
approach, because in complex AMR applications
the boxes typically have different
sizes,resulting in load imbalance among threads, and because there will 
typically be more threads than grids on many-core architectures 
unless many small grids are used.  Increasing the amount of parallelism 
by using many small grids in an AMR framework such as BoxLib is not 
recommended.  It increases the size of the AMR metadata,  increases the cost 
of performing box-box intersections needed to fill ghost cells and copy data between AMR levels, 
and increases the actual computation because of the increased volume of ghost cells.
All of the benefits one would achieve with many small grids are achieved by
tiling the grids instead.

Besides providing a backward compatible tiling capability and a
coarse-grained tile-level threading approach, the new version of
BoxLib also includes many performance improvements that
are made possible by applying the tiling concept.  One such example is
the operation of filling ghost cells, which involves both local and remote
communications.  For data available on its own MPI process, simple
copy operations are performed, whereas to get data from a remote MPI
process, MPI calls must be made.  In BoxLib, MPI messages between
processes are aggregated by using buffers for MPI
send and receive calls to reduce communication overhead and to
increase the utility of network bandwidth.  Thus the remote communication also involves local
data movement.  These local data copy operations are usually considered cheap.
However, on a many-core architecture with hundreds of threads, they
could become a bottleneck if they are not threaded or have low parallel
efficiency, as expected from Amdahl's law \cite{amdahlslaw}.   In
Section~\ref{sec:perf-heat}, we will present such an example and
demonstrate the improvement from using tiling.

The user function called inside the MFIter loop often dynamically
allocates its own temporary arrays. In a multi-threaded environment,
this often results in lock contention and sometimes translation
lookaside buffer (TLB) misses or even page faults.  One way of
avoiding these issues is using Fortran automatic arrays, which are
usually put on the stack instead of the heap by default (see
\cite{smc-pearl}) for an example of this approach).  Another approach
we have implemented in BoxLib is creating thread private memory arenas
and allocating dynamic arrays from the arenas.  One advantage of the
latter approach is that it provides the flexibility of allocating the
high-bandwidth on-package memory available on the next generation of
chips (e.g., Intel Knights Landing).

Finally we note that tiling is not always desired or better.  The
traditional fine-grained loop-level threading approach coupled with dynamic scheduling is
more appropriate for work with unbalanced loads, such as calculating
chemical reactions in cells by an implicit solver such as VODE \cite{vode}.

\section{Performance}
\label{sec:performance}

In this section, we present several results demonstrating
the performance benefits of tiling in BoxLib.  In the first test we
consider an explicit heat equation solver built with the 
BoxLib \Cplusplus framework, and in the second test we use a 
minimalist version of SMC \cite{SMC} built with
the BoxLib Fortran framework.  Unless stated otherwise, we define speedup 
as the ratio of the run time to the corresponding time in the 1-thread untiled run. 
Note that MPI is not used in these tests so that we can accurately measure the 
on-node performance impact of tiling.

\subsection{Heat Equation Solver}
\label{sec:perf-heat}

In our tests of tiling in the heat equation solver, the code took 
1000 time steps on a single level using a forward Euler method with 
second-order centered differences for the spatial derivatives 
\footnote{The source code of the test is available at 
{\tt Tutorials/Tiling\_Heat\_C} in BoxLib.}.  
The domain contained a single grid with $128^3$ cells,  
and periodic boundaries were assumed in all three dimensions.  
The computation of second derivatives in the heat equation kernel is
performed in two steps with three separate loops for the flux
components and another loop for the divergence of the flux.  Although
these loops in this memory-bound situation can be fused to improve
cache locality, this implementation is a typical pattern for
application developers without specific computer architecture
knowledge.  This test was intentionally simple so that the impact of
tiling can be clearly identified.

The first test was a series of non-threaded runs carried out on a single core of
Edison at the National Energy Research Scientific Computing Center
(NERSC), using either the default Gnu or Intel compiler. 
In these runs we varied the tile size and found that even with no parallelism,
tiling can significantly improve the performance due to the smaller working set size
(Table~\ref{tab:heat-tilesize}).  In this serial test, 
a factor of 1.8 speedup (Intel) and 
a factor of 3.4 speedup (Gnu),
due entirely to logical tiling, were achieved.

\begin{table}[htbp]
\caption{Impact of tiling on the performance of the heat equation kernel on
  the Edison machine at NERSC using the Gnu and Intel compilers.  
  We show the run time and speedup of the computational kernel
  for various tile sizes. The tile size refers to the
  number of cells in a logical tile in the $x$, $y$ and $z$-direction,
  respectively. \label{tab:heat-tilesize} } 
\begin{center}\footnotesize
\renewcommand{\arraystretch}{1.3}
\begin{tabular}{|l||c|c||c|c|}\hline
  {{\bf Tile Size}} & {{\bf Time - Gnu (s)}} & {{\bf Speedup - Gnu}} & {{\bf Time - Intel (s)}} & {{\bf Speedup - Intel}}\\
  \hline
 $128 \times 4 \times 4$   &  8.5 & 3.4 & 8.7   & 1.8\\
 $128 \times 8 \times 8$   &  9.0 & 3.2 & 9.6   & 1.6\\
 $128 \times 16 \times 16$ &  9.6 & 3.0 & 10.5  & 1.5\\
 $128 \times 32 \times 32$ & 23.7 & 1.2 & 10.4  & 1.5\\
 $128 \times 64 \times 64$ & 24.4 & 1.2 & 10.9 & 1.4\\
 no tiling                 & 28.6 & --  & 15.5 & --    \\
  \hline
\end{tabular}
\end{center}
\end{table}

\begin{figure}[h]
\centering
\includegraphics[width=3.5in]{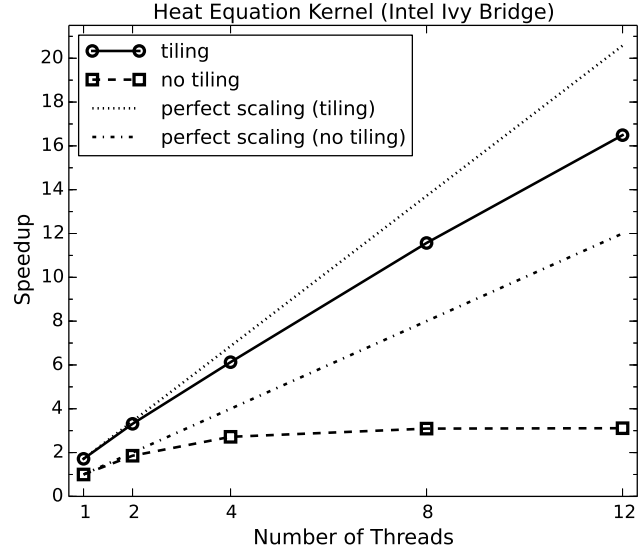}
\caption{\label{fig:heat-ivybridge} Heat equation kernel speedup on
  12-core Intel Ivy Bridge.  We compare the results of tiled and
  untiled runs. A 16.5x speedup was obtained in the 12-thread tiled
  run. }
\end{figure}

We next explored the parallel performance of the tiled heat equation
solver.  We performed a strong scaling study on Edison, which has 
12-core Intel Ivy Bridge processors, and used OpenMP for threading.
Figure~\ref{fig:heat-ivybridge} compares the tiled runs with untiled
runs where a tile size of $128 \times 4 \times 4$ cells was used for
all of the tiled runs.  Two different threading approaches were used; 
the tiled runs used the coarse-grained tile-level threading approach
as described in Section~\ref{sec:boxlib1}, whereas the
untiled runs used the fine-grained loop-level threading approach
described in Section~\ref{sec:boxlib2}.  The speedup shown in 
Figure~\ref{fig:heat-ivybridge} is for the computational kernel only. 
In this figure the tile-level threading clearly demonstrates much better 
scaling behavior than the loop-level approach.  The 12-thread
tiled run was more than 5.3 times faster than the 12-thread untiled
run, and it had a 16.5x speedup over the single thread untiled run.
For the tiled runs, a 9.6x speedup over the single thread tiled run
was obtained with 12 threads.  We note that perfect strong scaling
within a socket is difficult to achieve because Edison has 30 MB L3
cache shared among 12 cores on a processor.

\begin{figure}[h]
\centering
\includegraphics[width=3.5in]{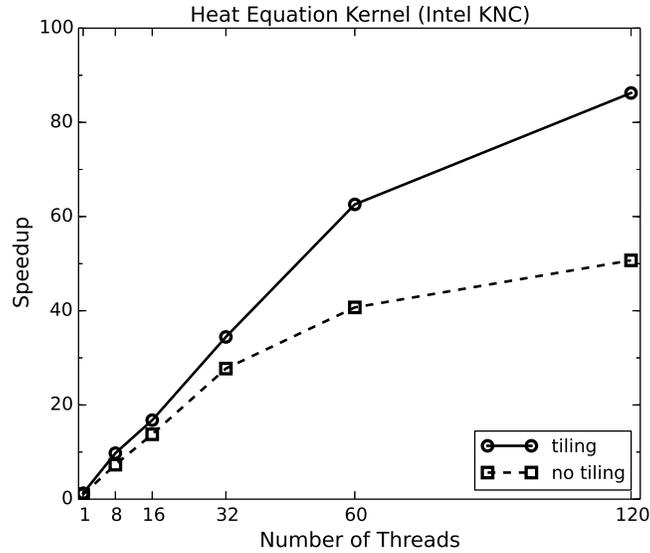}
\caption{\label{fig:heat-knc} Heat equation kernel speedup on 60-core
  Intel KNC.  We compare the results of tiled and untiled runs.  A
  speedup of 86x was obtained for the 120-thread tiled run. }
\end{figure}

A similar strong scaling study of the heat equation was carried out
on a single 60-core Intel Knights Corner processor on NERSC's Babbage
computer.  We again observed the performance benefit of the tiling approach with
coarse-grained tile-level parallelism over the non-tiling
approach with fine-grained loop-level parallelism.
Figure~\ref{fig:heat-knc} shows the speedup of the computational kernel
for untiled and tiled runs.  The 120-thread tiled
run had a speedup of 69x and 86x with respect to the 1-thread
tiled and non-tiled runs, respectively.  The 60-core Intel KNC
processors on Babbage are capable of spawning 4 hardware threads per
core.  The results using 180 and 240 threads are not shown in the
figure because no speedup of the computational kernel was obtained
beyond 120 threads.  In fact the 240-thread run was about 20\% slower
than the 120-thread run.  We note that no effort was spent on
optimizing the kernel for KNC.

\begin{figure}[h]
\centering
\includegraphics[width=3.5in]{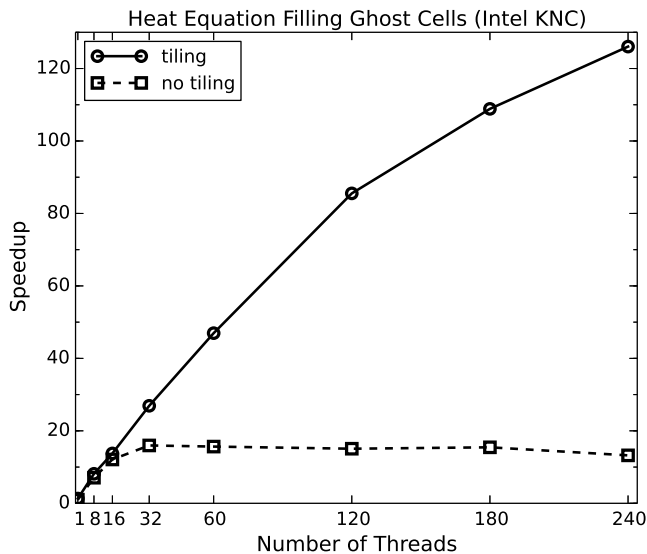}
\caption{\label{fig:heat-fpb} Speedup in filling ghost cells in the heat equation solver
  on 60-core Intel KNC.  We compare the results of tiled and untiled runs.  A
  speedup of 126x was obtained for the 240-thread tiled run. }
\end{figure}

\begin{figure}[h]
\centering
\includegraphics[width=3.5in]{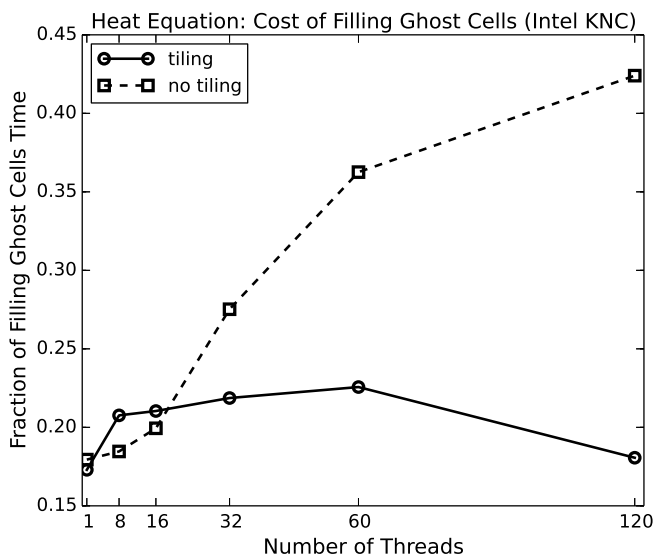}
\caption{\label{fig:heat-time} Heat equation: cost of filling ghost
  cells time on 60-core Intel KNC.  We show the ratio of the time spent 
  filling ghost cells to the sum of filling ghost cells and kernel
  times.  Note that a significant fraction of the time was spent 
  filling ghost cells for the untiled runs.}
\end{figure}

In the strong scaling study described above, we also measured the time to 
fill ghost cells at the periodic boundaries.
Note again that MPI was not used and the data movement was completely local.  
The scaling results for just filling the ghost cells are shown in
Figure~\ref{fig:heat-fpb}.  The 240-thread tiled run had a speedup of
96x and 126x in filling ghost cells with respect to the 1-thread
tiled and non-tiled runs, respectively.  Figure~\ref{fig:heat-time}
shows the fraction of time spent filling ghost cells defined as
fill time / (fill time + kernel time).  It is a striking
reminder of Amdahl's law \cite{amdahlslaw}
that the 120-thread non-tiled run spent more than 40\% of the time filling
ghost cells.  

\subsection{SMC}
\label{sec:perf-smc}

The SMC code solves the multicomponent, reacting, compressible Navier-Stokes
equations using a mixture model for species diffusion \cite{SMC}.  
The spatial derivatives are computed with 8th-order stencils that
span 9 cells in each coordinate direction. 
Here we use a minimalist version of SMC for
performance testing \footnote{The source code of this minimalist
version of SMC is available at {\tt MiniApps/SMC} in BoxLib.}.  
In this minimalist version, Sutherland's viscosity law is adopted, a constant Schmidt
number is used for mass diffusivity, and a constant Prandtl number is
used to compute thermal diffusivity.  The runs considered here used a 9-species
$\mathrm{H_2}/\mathrm{O_2}$ reaction mechanism \cite{LiDryer} 
and a 3rd-order Runge-Kutta method was used for time integration.  
We have previously demonstrated excellent scaling behavior of manually tiled
SMC \cite{SMC}.  The version of SMC described here is based on the new
version of BoxLib described in this paper.

\begin{figure}
\centering
\includegraphics[width=3.5in]{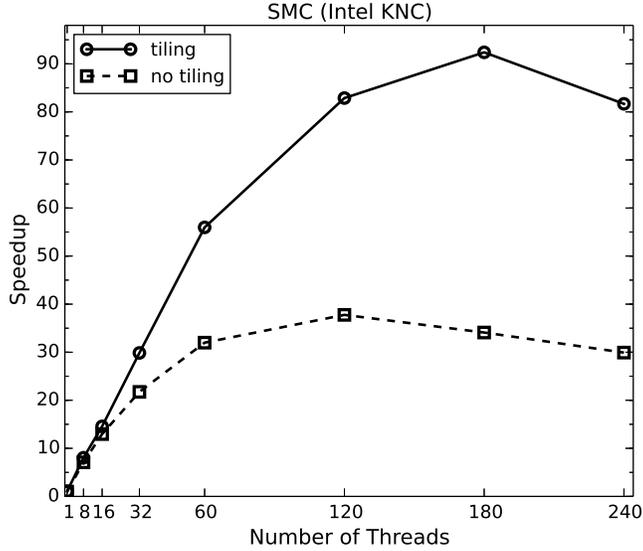}
\caption{\label{fig:smc-knc} SMC speedup on 60-core Intel KNC.  We
  show the speedup of the whole run excluding the one-time
  initialization.  The 180-thread tiled run had a 92x speedup over the
single thread untiled run.}
\end{figure}

A series of SMC runs were carried out on a single 60-core processor on Babbage.
As with the heat equation solver, the domain contained a single 
grid of $128^3$ cells and periodic boundaries were assumed in all three
dimensions.  The tile size was again chosen to be $128 \times 4 \times 4$
in the $x$, $y$ and $z$-direction, respectively.  For comparison, we have
performed both tiled and untiled runs.  As above, 
the tiled runs used the coarse-grained tile-level threading approach
whereas the untiled runs used the fine-grained loop-level threading approach.
The scaling results are shown
in Figure~\ref{fig:smc-knc}.  A speedup of 86x was obtained for the
180-thread tiled run over the single thread tiled run, similar to the
result obtained with the manual tiling.  Moreover, the best tiled
run was 2.4 times faster than the best untiled run, and a 92x
speedup was obtained for the 180-thread tiled run over the single
thread untiled run.  The results demonstrate clearly that excellent
performance can be obtained using BoxLib with tiling on many-core
architectures with more than 100 threads.

\section{Future Work} 
\label{sec:regional}

Future work for the BoxLib framework includes integrating the TiDA library currently under
development into the BoxLib framework
in order to enable regional tiling as well as logical tiling.  Regional tiling addresses potential
NUMA issues that are likely to become even more prevalent on next-generation architectures.
Such architectures enable a thread placed on a NUMA domain to access 
remote memory of another domain at a cost of higher latency and lower bandwidth, 
often referred to as NUMA effects.  One can address NUMA effects by either placing data in the memory 
local to where most computations will be carried out, or by moving computations to the NUMA domain that holds the data.
However, neither memory management nor thread management is a trivial task for the application developer.
The approach currently adopted in BoxLib is to place one MPI process per NUMA node 
and use a threaded model such as OpenMP within a NUMA node.  
However, such a model results in higher overhead for inter-process synchronization (e.g. barrier and reduction),
due to more MPI processes, than a model in which a single MPI process spans multiple NUMA nodes.
In addition, MPI often requires additional memory to buffer messages at both the application and system levels.
In future architectures we expect to see a significant reduction in memory capacity per core and a substantial increase in the number of 
NUMA domains per compute node \cite{padal-report}.   Regional tiling allows one to respect NUMA effects while operating
with one MPI process per compute node.

\begin{figure}[h]
\centering
\includegraphics[width=3.5in]{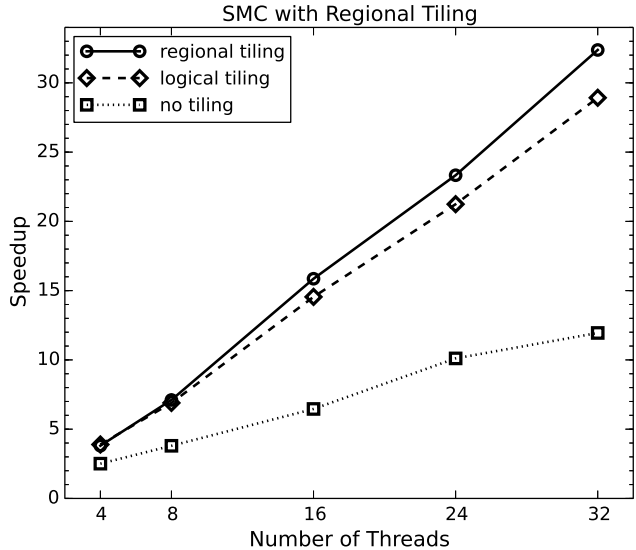}
\caption{\label{fig:smc-tida} SMC with regional tiling on Trestles}
\end{figure}

The TiDA library~\cite{TiDA,TiDA-ISC} will provide developers with NUMA-aware tiling and
layout abstractions at the application level and hides details of how
regions and tiles are constructed, mapped and executed on various hardware platforms.  
The layout abstraction manages the memory allocation for regions and keeps track of their location on physical partitions of the memory.
In Figure~\ref{fig:smc-tida} we show a performance comparison between prototype logical and regional tiling implementations 
of SMC.  In this example, we conducted a strong scaling study on up to 8 NUMA domains (i.e. 32 cores) 
of a compute node on Trestles, a system at San Diego Supercomputer Center.
It can be seen that when only one NUMA node (4 cores) is used, the performance of both implementations matches.
However, with more NUMA domains regional tiling overtakes logical tiling, and the performance gap expands as the number of NUMA domains increases.  
The reason is that regional tiling reduces NUMA effects by allocating each region on a different NUMA node. 
In particular, with logical tiling a tile may access remote memory if a neighbor resides in a different NUMA domain.
Latency costs will become significant when there are many remote memory accesses on small amounts of data.
With regional tiling, tiles at the boundary exchange ghost cells, realizing higher bandwidth and amortizing the latency overhead.
In this protoype version of the SMC code, regional tiling achieves 32x the performance over one thread on 32 cores.

\section{Summary}
\label{sec:summary}

In order to address the performance challenges that will accompany 
next generation node architectures based on many-core processors with NUMA domains,
we have implemented logical tiling in BoxLib, a block-structured adaptive mesh refinement 
software framework.  
In this paper we have described how the tiling constructs
are embedded in the iterator constructs in BoxLib so that BoxLib-based applications 
can easily realize the expected serial and parallel performance gains 
without extra effort on the part of the application developer.
We also discussed a path forward to enable
future versions of BoxLib to take advantage of NUMA-aware optimizations
using the TiDA portable library.

\bibliographystyle{siam}
\bibliography{ws}

\begin{thebibliography}{10}

\bibitem{BoxLib}
Center for Computational Sciences and Engineering, Lawrence Berkeley National
  Laboratory, {\emph{Box{L}ib}}, software available at
  \url{https://ccse.lbl.gov/BoxLib/}.

\bibitem{Nyx}
{\sc A.S. Almgren, J.B. Bell, M.J. Lijewski, Z.~Lukic, and E.~Van Andel}, {\em
  Nyx: A massively parallel {AMR} code for computational cosmology}, APJ, 765
  (2013), p.~39.

\bibitem{CASTRO_I}
{\sc A.~S. {Almgren}, V.~E. {Beckner}, J.~B. {Bell}, M.~S. {Day}, L.~H.
  {Howell}, C.~C. {Joggerst}, M.~J. {Lijewski}, A.~{Nonaka}, M.~{Singer}, and
  M.~{Zingale}}, {\em {CASTRO: A New Compressible Astrophysical Solver. I.
  Hydrodynamics and Self-gravity}}, ApJ, 715 (2010), pp.~1221--1238.

\bibitem{ABNZ:III}
{\sc A.~S. Almgren, J.~B. Bell, A.~Nonaka, and M.~Zingale}, {\em Low {Mach}
  number modeling of {Type Ia} supernovae. {III. Reactions}}, APJ, 684 (2008),
  pp.~449--470.
\newblock paper III.

\bibitem{ABRZ:I}
{\sc A.~S. Almgren, J.~B. Bell, C.~A. Rendleman, and M.~Zingale}, {\em Low
  {Mach} number modeling of {Type Ia} supernovae. {I. Hydrodynamics}}, APJ, 637
  (2006), pp.~922--936.
\newblock paper I.

\bibitem{ABRZ:II}
\leavevmode\vrule height 2pt depth -1.6pt width 23pt, {\em Low {Mach} number
  modeling of {Type Ia} supernovae. {II. Energy Evolution}}, APJ, 649 (2006),
  pp.~927--938.
\newblock paper II.

\bibitem{amdahlslaw}
{\sc Gene~M Amdahl}, {\em Validity of the single processor approach to
  achieving large scale computing capabilities}, in Proceedings of the April
  18-20, 1967, spring joint computer conference, ACM, 1967, pp.~483--485.

\bibitem{LMC3}
{\sc A.J. Aspden, M.~S. Day, and J.~B. Bell}, {\em Turbulence-chemistry
  interaction in lean premixed hydrogen combustion}, Proc. Comb. Inst., 35
  (2014), pp.~1321--1329.

\bibitem{LMC2}
{\sc J.B. Bell, M.S. Day, and M.J. Lijewski}, {\em Simulation of nitrogen
  emissions in a premixed hydrogen flame stabilized on a low swirl burner},
  Proc. Comb. Inst., 34 (2013), pp.~1173--1182.

\bibitem{HTA}
{\sc Ganesh Bikshandi, Jia Guo, Daniel Hoeflinger, Gheorghe Almasi, Basilio~B.
  Fraguela, Mar\'{\i}a~J. Garzar\'{a}n, David Padua, and \~Christoph von
  Praun}, {\em Programming for parallelism and locality with hierarchically
  tiled arrays}, in Proceedings of the eleventh ACM SIGPLAN symposium on
  Principles and practice of parallel programming, PPoPP '06, New York, NY,
  USA, 2006, ACM, pp.~48--57.

\bibitem{vode}
{\sc P.~Brown, G.~Byrne, and A.~Hindmarsh}, {\em Vode: A variable-coefficient
  ode solver}, SIAM Journal on Scientific and Statistical Computing, 10 (1989),
  pp.~1038--1051.

\bibitem{Carr:1992:CBN}
{\sc S.~Carr and K.~Kennedy}, {\em Compiler blockability of numerical
  algorithms}, in Proceedings of the 1992 ACM/IEEE conference on
  Supercomputing, Supercomputing '92, Los Alamitos, CA, USA, 1992, IEEE
  Computer Society Press, pp.~114--124.

\bibitem{DayBell:2000}
{\sc M.~S. Day and J.~B. Bell}, {\em Numerical simulation of laminar reacting
  flows with complex chemistry}, Combust. Theory Modelling, 4 (2000),
  pp.~535--556.

\bibitem{CASTROmoist}
{\sc M.~Duarte, A.~S. Almgren, K.~Balakrishnan, J.~B. Bell, and D.~M. Romps},
  {\em A numerical study of methods for moist atmospheric flows: {C}ompressible
  equations}, Mon. Wea. Rev., 142 (2014), pp.~4269--4283.

\bibitem{MAESTROmoist}
{\sc M.~Duarte, A.~S. Almgren, and J.~B. Bell}, {\em A low mach number model
  for moist atmospheric flows}, J. Atmos. Sci., 72 (2015), pp.~1605--1620.

\bibitem{Dubeyetal:2014}
{\sc A.~Dubey, A.~Almgren, J.~Bell, M.~Berzins, S.~Brandt, G.~Bryan, D.~Graves
  P.~Colella, M.~Lijewski, F.~Loffler, B.~O'Shea, E.~Schnetter, B.~Van
  Straalen, and K.~Weide}, {\em A survey of high level frameworks in
  block-structured adaptive mesh refinement packages}, Journal of Parallel and
  Distributed Computing, 74 (2014), pp.~3217--3227.

\bibitem{SMC}
{\sc Matthew Emmett, Weiqun Zhang, and John~B. Bell}, {\em High-order
  algorithms for compressible reacting flow with complex chemistry}, Combustion
  Theory and Modelling, 18 (2014), pp.~361--387.

\bibitem{Goumas:2004:APC}
{\sc Georgios~Goumas et~al.}, {\em Automatic parallel code generation for tiled
  nested loops}, in Proceedings of the 2004 ACM symposium on Applied computing,
  SAC '04.

\bibitem{FLASH}
{\sc B.~{Fryxell}, K.~{Olson}, P.~{Ricker}, F.~X. {Timmes}, M.~{Zingale}, D.~Q.
  {Lamb}, P.~{MacNeice}, R.~{Rosner}, J.~W. {Truran}, and H.~{Tufo}}, {\em
  {FLASH}: An adaptive mesh hydrodynamics code for modeling astrophysical
  thermonuclear flashes}, Astrophysical Journal Supplement, 131 (2000),
  pp.~273--334.

\bibitem{LiDryer}
{\sc Juan Li, Zhenwei Zhao, Andrei Kazakov, and Frederick~L. Dryer}, {\em An
  updated comprehensive kinetic model of hydrogen combustion}, International
  Journal of Chemical Kinetics, 36 (2004), pp.~566--575.

\bibitem{Nyx2}
{\sc Zarija Lukic, Casey Stark, Peter Nugent, Martin White, Avery Meiksin, and
  Ann Almgren}, {\em The lyman-alpha forest in optically-thin hydrodynamical
  simulations}, MNRAS, 446 (2015), pp.~3697--3724.

\bibitem{Malone_etal:2014}
{\sc C.~M. Malone, A.~Nonaka, S.~E. Woosley, A.~S. Almgren, J.~B. Bell,
  S.~Dong, and M.~Zingale}, {\em The deflagration stage of chandrasekhar mass
  models for type ia supernovae. i. early evolution}, The Astrophysical
  Journal, 782 (2014), p.~11.

\bibitem{MAESTRO:Multilevel}
{\sc A.~Nonaka, A.~S. Almgren, J.~B. Bell, M.~J. Lijewski, C.~M. Malone, and
  M.~Zingale}, {\em {\tt MAESTRO}:an adaptive low mach number hydrodynamics
  algorithm for stellar flows}, APJ:sup, 188 (2010), pp.~358--383.
\newblock paper V.

\bibitem{Pau:2012}
{\sc George Shu~Heng Pau, John~B Bell, Ann~S Almgren, Kirsten~M Fagnan, and
  Michael~J Lijewski}, {\em An adaptive mesh refinement algorithm for
  compressible two-phase flow in porous media}, Computational Geosciences, 16
  (2012), pp.~577--592.

\bibitem{Rivera:2000}
{\sc Gabriel Rivera and Chau-Wen Tseng}, {\em Tiling optimizations for 3d
  scientific computations}, in Proceedings of the 2000 ACM/IEEE Conference on
  Supercomputing, SC '00, Washington, DC, USA, 2000, IEEE Computer Society.

\bibitem{Rivera:2000:tiling}
{\sc Gabriel Rivera and Chau~Wen Tseng}, {\em Tiling optimizations for 3d
  scientific computations}, in Proceedings of the 2000 ACM/IEEE conference on
  Supercomputing, Supercomputing '00, Washington, DC, USA, 2000, IEEE Computer
  Society.

\bibitem{Song:1999:NTT}
{\sc Yonghong Song and Zhiyuan Li}, {\em New tiling techniques to improve cache
  temporal locality}, SIGPLAN Not., 34 (1999), pp.~215--228.

\bibitem{ICS11:Mint}
{\sc Didem Unat, Xing Cai, and Scott~B. Baden}, {\em {Mint: realizing CUDA
  performance in 3D stencil methods with annotated C}}, in Proceedings of the
  international conference on Supercomputing, ICS '11, New York, NY, USA, 2011,
  ACM, pp.~214--224.

\bibitem{TiDA}
{\sc Didem Unat, Cy~Chan, Weiqun Zhang, John Bell, and John Shalf}, {\em Tiling
  as a durable abstraction for parallelism and data locality}, Workshop on
  Domain-Specific Languages and High-Level Frameworks for High Performance
  Computing,  (November 18, 2013).

\bibitem{TiDA-ISC}
{\sc Didem Unat, Tan Nguyen, Weiqun Zhang, Muhammed~Nufail Farooqi, Burak
  Bastem, George Michelogiannakis, Ann Almgren, and John Shalf}, {\em Tida:
  High-level programming abstractions for data locality management}, in
  Supercomputing, Lecture Notes in Computer Science, Springer Berlin
  Heidelberg, 2016.

\bibitem{padal-report}
{\sc Didem Unat, John Shalf, Torsten Hoefler, Thomas Schulthess, Anshu~Dubey
  (Editors), et~al.}, {\em {Programming Abstractions for Data Locality}}, tech.
  report, 04 2014.

\bibitem{smc-pearl}
{\sc Antonio Valles and Weiqun Zhang}, {\em Optimizing for reacting
  navier-stokes equations}, in High Performance Parallelism Pearls Volume One:
  Multicore and Many-core Programming Approaches, James Reinders and James
  Jeffers, eds., Morgan Kaufmann, 2014.

\bibitem{Wolfe:1989}
{\sc M.~Wolfe}, {\em More iteration space tiling}, in Proceedings of the 1989
  ACM/IEEE conference on Supercomputing, Supercomputing '89, New York, NY, USA,
  1989, ACM, pp.~655--664.

\bibitem{CASTRO_II}
{\sc W.~Zhang, L.~Howell, A.~Almgren, A.~Burrows, and J.~Bell}, {\em Castro: A
  new compressible astrophysical solver. ii. gray radiation hydrodynamics},
  APJ:sup, 196 (2011), p.~20.

\bibitem{CASTRO_III}
{\sc W.~Zhang, L.~Howell, A.~Almgren, A.~Burrows, J.~Dolence, and J.~Bell},
  {\em Castro: A new compressible astrophysical solver. iii. multigroup
  radiation hydrodynamics}, APJ:sup, 204 (2013), p.~7.

\bibitem{ZABNW:IV}
{\sc M.~Zingale, A.~S. Almgren, J.~B. Bell, A.~Nonaka, and S.~E. Woosley}, {\em
  Low mach number modeling of {T}ype {Ia} supernovae. {IV}. {W}hite dwarf
  convection}, APJ, 704 (2009), pp.~196--210.

\end{thebibliography}

\end{document}